\documentclass[conference]{IEEEtran}

\IEEEoverridecommandlockouts

\usepackage{amsmath,amsthm,amssymb}
\usepackage{algorithm}
\usepackage[noend]{algpseudocode}
\usepackage{algorithmicx}
\usepackage{verbatim}

\usepackage{pgfplots}
\pgfplotsset{compat=newest}
\pgfplotsset{plot coordinates/math parser=false}
\newlength\figureheight
\newlength\figurewidth
\newlength\figureheightB
\newlength\figurewidthB
\newlength\figureheightC
\newlength\figurewidthC
\newlength\figureheightD
\newlength\figurewidthD
\pgfplotsset{major grid style={dashed,line width=0.5pt}}

\graphicspath{{./}}

\usepackage{hyperref}
\hypersetup{
  colorlinks   = true,    
  urlcolor     = blue,    
  linkcolor    = blue,    
  citecolor    = red      
}

\begin{document}

\title{Cooperative Network Node Positioning Techniques Using Underwater Radio Communications}

\author{\IEEEauthorblockN{Javier~Zazo,
 Santiago~Zazo,\\
 Sergio~Valcarcel~Macua,
 Marina P\'{e}rez}
 \vspace{1em}
\IEEEauthorblockA{
Universidad Polit\'ecnica de Madrid,\\
ETSIT, Madrid, 28040, Spain.\\
\{javier.zazo.ruiz,santiago.zazo\}@upm.es}
\thanks{This work was supported in part by the Spanish Ministry of Science and
Innovation under the grant TEC2013-46011-C3-1-R (UnderWorld), the
COMONSENS Network of Excellence TEC2015-69648-REDC and by an FPU
doctoral grant to Javier Zazo.}
\and
\IEEEauthorblockN{Iv\'{a}n P\'{e}rez-\'{A}lvarez,\\ Eugenio Jim\'{e}nez}
 \vspace{1em}
\IEEEauthorblockA{
Universidad de Las Palmas de Gran Canaria,\\
IDeTIC, Las Palmas, 35017, Spain.\\
ivan.perez@ulpgc.es}
\and
\IEEEauthorblockN{Laura Cardona, \\Eduardo Quevedo}
 \vspace{1em}
\IEEEauthorblockA{
Oceanic Platform of the\\ Canary
Islands (PLOCAN),\\
Telde, 35200, Spain.
}
}



\maketitle

%
\begin{abstract}
We analyze the problem of localization algorithms for underwater sensor networks.
We first characterize the underwater channel for radio communications and adjust a linear model with measurements of real transmissions.
We propose an algorithm where the sensor nodes collaboratively estimate their unknown positions in the network.
In this setting, we assume low connectivity of the nodes, low data rates, and non-zero probability of lost packets in the transmission.
Finally, we consider the problem of a node estimating it's position in underwater navigation.
We also provide simulations illustrating the previous proposals.
\end{abstract}
%


\IEEEpeerreviewmaketitle

%
\section{Introduction}
%
Underwater sensor networks result as a promising technology to deal with applications that require off-shore monitoring and inexpensive deployment. 
Environmental quality control, seismic analysis, oilfield monitoring and sea robotics are some applications where underwater networks may provide a viable solution to  these open problems~\cite{Heidemann2012}.
The wireless underwater communication presents, however, a challenging task that needs to be adapted to different applications in consideration.

There are three main physical access channels that can be considered for underwater wireless communications, being acoustic, optical and radio.
Each of them presents their own benefits and drawbacks, and can be used in different scenarios~\cite{Che2010,Palmeiro2011}.
For instance, acoustic communications are valid for medium range distances--several km~\cite{Farr2010}--but has limited bandwidth, poor performance in shallow water and has impact on marine life~\cite{Parsons2008,Jepson2003}.
Optical communications, on the contrary, present ultra-high bandwidth over short distances but are susceptible to turbidy, particles or marine fouling. Furthermore, ambient light is another adverse effect which does not make this technology suitable for shallow waters~\cite{Lanbo2008}.

Radio frequency transmissions--in the range of KHz or MHz--become a worth-studying option to establish communication between the nodes in shallow waters, such as in lakes, bays, harbors and areas close to the sea shores.
Although the water medium still has a strong attenuation in these frequencies, communication is possible in the range of meters. 
Companion papers~\cite{Jimenez2016} and~\cite{Mena2016} provide specific measurements and describe the procedure for transmitting pure sinusoidal signals at different frequencies and distances.
The results obtained in these publications establish a descriptive framework, where a model of the physical layer is derived, and can be used for the analysis of higher end applications.

The next challenge once the physical layer has been modeled accounting to real measurements, as described in~\cite{Jimenez2016,Mena2016}, is to develop the protocols to effectively establish a communication channel between two different nodes that operate in the underwater sensor network.
The specific details to analyze the performance of such network through simulations is considered in our companion paper~\cite{Valcarcel2016}.
In such reference we consider the model of the channel, the modulation parameters, the MAC protocol and the retransmission strategy under packet transference errors.
With such considerations we present an analysis tool to effectively study the viability of different network configurations in shallow waters.

Our contribution in this publication extends~\cite{Jimenez2016,Mena2016} and~\cite{Valcarcel2016} and focuses on the application layer.
We consider the problem of self-estimating the network nodes position after deployment, as well as discussing localization methods to determine the position of an external node, such as an Underwater Unmanned Vehicle (UUV).
The specific conditions of underwater communications require the use of improved algorithms that account for low data rates, low connectivity between the nodes, and do not present significant performance loss if packets are lost in the transmission.
In this paper we use state of the art localization methods and adapt them for the specific restrictions of underwater radio networks.

In Section~\ref{sec:network-description} we describe the channel model and communication protocols.
In Section~\ref{sec:network-positions} we propose an algorithm for the nodes to determine their own position and adapt it for low communication rates.
We also study its performance under transmitted packet losses.
In Section~\ref{sec:navigation-aid} we consider the application of the network nodes aiding in the navigation of a UUV.
We consider an algorithm that works under minimal communication and is suitable for the underwater channel.
Finally, in Section~\ref{sec:simulations} we present performance simulations of the previous proposals.

%
\section{Network Description}
\label{sec:network-description}
%
We consider an underwater network formed of $N$ nodes and we refer to the the set of nodes with $\mathcal{N}=\{ 1,\ldots,N \}$.
These nodes communicate in a wireless environment using loop antennas as presented in~\cite{Jimenez2016,Mena2016}.
The specific signal strengths measured at fixed distances between receiver and transmitter are presented in Figure~\ref{fig:measures}, where it can be observed that the medium presents lowest attenuation at the lower frequencies.
The transmitted power was set to 20 dBm on every curve in all measurements.
Note also that the attenuation increases exponentially in higher frequencies, and because of that, the overall bandwidth must be narrow.
For the previous reasons, and as a design choice, we established the communication frequency at 33~KHz and bandwidth of 6~KHz.

The previous channel limitations require a simple modulation that can work under high attenuation on a non-flat channel.
Our design choice is to use a binary frequency-shift keying (BFSK) with first carrier at 30~KHz and second carrier at 36~KHz.
The choice of these parameters allow a communication rate of about 3~kbps.
We can represent the channel attenuation as a function of distance for the BFSK frequencies and propose a model to calculate distances over measured powers.
These measurements are represented in Figure~\ref{fig:dist-measurements}, where points indicate power measurements, and the line corresponds to a linear model of form:
\begin{equation}
	g = a d + b + \varepsilon.
	\label{eq:linear-model}
\end{equation}
In the model $g$ represents the gain of the channel; $d$ is the given distance between nodes; $a$ and $b$ are variables to be determined by the linear model; and  $\varepsilon$ is an additive Gaussian noise of the measurements with zero mean.
Given the previous model, the maximum likelihood (ML) estimator is given by the solution to the following problem:
\begin{equation}
	\min_x \Vert A x -y\Vert^2
\end{equation}
where $x=[a,b]^T$, $y=[g_1,\ldots,g_n]$ represent the measured gains at distances $[d_1,d_2,\cdots, d_n]$ and
\begin{equation}
	A^T =
	\begin{pmatrix}
		d_1 & d_2 & \cdots & d_n \\
		1 & 1 & \cdots & 1
	\end{pmatrix}.
\end{equation}
For the particular measurements our collaborators performed in~\cite{Jimenez2016,Mena2016}, our linear model becomes: $a\approx-8.5~\rm{dB/m}$ and $b\approx -54.85~\rm{dB}$.
Regarding random variable~$\varepsilon$ we can estimate its variance with the ML estimator as follows:
\begin{equation}
	\sigma^2_{\varepsilon} = \frac{1}{n}\Vert Ax^\ast-y\Vert^2=1.15~\rm{dB}
	\label{eq:noice-variance}
\end{equation}
where $n$ indicates the number of measurements performed.

\begin{figure}
	\centering
    \input{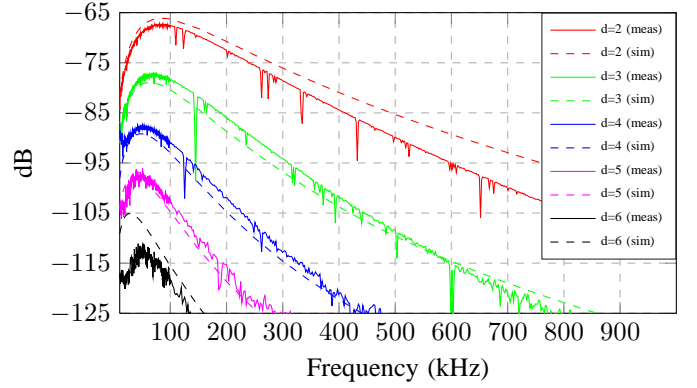}
	\caption{Attenuation between two horizontal ten-turns loops placed on seabed.}
	\label{fig:measures}
\end{figure}
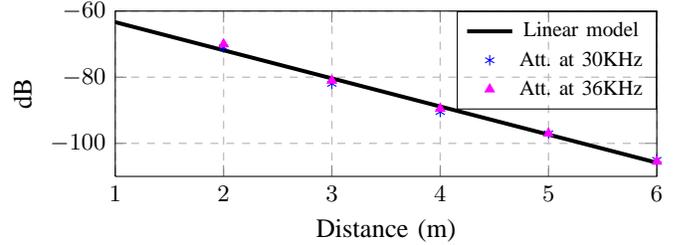
\begin{figure}
	\centering
%
%
\definecolor{mycolor1}{rgb}{1.00000,0.00000,1.00000}%
\begin{tikzpicture}

\begin{axis}[%
width=\figurewidth,
height=\figureheight,
at={(0.758333in,0.48125in)},
scale only axis,
xmin=1,
xmax=6,
xlabel={Distance (m)},
xmajorgrids,
ymin=-110,
ymax=-60,
ylabel={dB},
ymajorgrids,
ticklabel style = {font=\small},
legend style={at={(1,1)},anchor=north east,font=\footnotesize,fill=none}
]
\addplot [color=black,solid,line width=1.5pt]
  table[row sep=crcr]{%
1	-63.35\\
2	-71.85\\
3	-80.35\\
4	-88.85\\
5	-97.35\\
6	-105.85\\
};
\addlegendentry{Linear model};

\addplot [color=blue,only marks,mark=asterisk,mark options={solid}]
  table[row sep=crcr]{%
2	-71\\
3	-82\\
4	-90.5\\
5	-97\\
6	-105\\
};
\addlegendentry{Att. at 30KHz};

\addplot [color=mycolor1,only marks,mark=triangle*,mark options={solid,fill=mycolor1}]
  table[row sep=crcr]{%
2	-70\\
3	-81\\
4	-89.5\\
5	-97\\
6	-105.5\\
};
\addlegendentry{Att. at 36KHz};

\end{axis}
\end{tikzpicture}%
	\caption{Attenuation vs. distance for $f_1=30~KHz$ and $f_2=36~KHz$.}
	\label{fig:dist-measurements}
\end{figure}

Finally, we describe the networks communication protocol, also part of the designer choice.
Nodes access the channel medium through a CSMA/CA protocol and communication is established between two nodes in an exclusive manner.
After each transmission the receiver confirms correct reception through an acknowledgement in order to avoid lost data in the channel.
If the message is not correctly transmitted, retransmission is performed following an automatic repeat protocol.
We assume that the message is lost if correct transmission is not achieved in a prefixed number of attempts.
Both protocols help to avoid the problem of hidden nodes and to establish a reliable connection.
The simulated implementation of these protocols for underwater channel analyzing the number of lost packets and energy expenditure is considered in~\cite{Valcarcel2016}.
In the following we take into account such results to implement our positioning algorithms.

%
\section{Network Node's Self-positioning Estimation}
\label{sec:network-positions}
%
In this section we propose a cooperative localization algorithm for nodes which do not know their own position, but can communicate with their neighbors to estimate their own coordinates.
We refer to the set of nodes that have knowledge of their position as ``anchor nodes'' and denote them with $\mathcal{N}_a$.
Likewise, the set of nodes that do not know their position are denoted as $\mathcal{N}_u$.
We also have $\mathcal{N}_a\cup\mathcal{N}_u=\mathcal{N}$.
We refer to the nodes that are within radio reach of node $i$ as neighbours, and denote them with $\mathcal{N}^i$.
Note that the set of neighbours may include both nodes from $\mathcal{N}_a$ and $\mathcal{N}_u$.

The localization algorithm is performed once when the nodes come online as an initial calibration, and we analyze it's performance in terms of convergence and lost packets.
The optimization problem we propose is given by:
\begin{IEEEeqnarray}{c}
\begin{IEEEeqnarraybox}[][c]{r'l}
	\min_{x_i\in\mathbb{R}^2} &
	\sum_{i\in\mathcal{N}_u} \sum_{j\in \mathcal{N}^i} (d_{ij}^2-\Vert x_i-x_j\Vert^2)^2  =\\
	\min_{x_i\in\mathbb{R}^2} & \sum_{i\in\mathcal{N}_u} \sum_{j\in \mathcal{N}^i} d_{ij}^4+\Vert x_i-x_j\Vert^4-2d_{ij}^2\Vert x_i-x_j\Vert^2.
\end{IEEEeqnarraybox}
\label{eq:ms_problem}
\end{IEEEeqnarray}
Problem~\eqref{eq:ms_problem} is nonconvex and can be solved using algorithms of the family of \emph{successive convex approximations}, such as the one proposed in~\cite{Scutari2014}.
The methodology consists in an iterative procedure where a convex function is used to approximate the non-convex original objective in a local region around a temporary fixed point. 
The surrogate function retains the convex parts and linearizes the non-convex ones.
The convexification is performed around a temporary pivotal point which is updated in subsequent iteration steps.
Specifically, the convex surrogate function we use is given by
\begin{equation}
 	\tilde{U}(x,x^{k}) = \sum_{i\in\mathcal{N}_u} \sum_{j\in \mathcal{N}^i} d_{ij}^4+\Vert x_i-x_j\Vert^4-4d_{ij}^2(x^{k}_i-x^{k}_j)x_i
\end{equation}
where $x^k=(x^k_i)_{i\in\mathcal{N}_u}$ is the pivotal point which is updated at different iteration steps.

The complete set of instructions to solve problem~\eqref{eq:ms_problem} is described in Algorithm~\ref{alg:scutari}.
Note that every user needs to solve the following problem:
\begin{IEEEeqnarray}{r'l}
	\hat{x_i}(x^k)) = \min_{x_i} & \tilde{U}(x_i,x^{k})
\label{eq:local-opt-scutari}
\end{IEEEeqnarray}
where with a little abuse of notation we explicitly stated that~\eqref{eq:local-opt-scutari} is minimized only in $x_i$ because the algorithm is solved in a distributed manner.
Problem~\eqref{eq:local-opt-scutari} is convex and can be solved using any convex optimization solver, or more efficiently, directly finding the solution to the KKT system of equations~\cite{Boyd2004}.
Moreover, and in order to solve~\eqref{eq:local-opt-scutari}, nodes only need to exchange their own estimates with their neighbours.
This makes Algorithm~\ref{alg:scutari} very convenient for distributed implementation.
Finally, note that convergence to a global minimum is not guaranteed since the problem is non-convex.

\begin{algorithm}
\begin{algorithmic}[1]
	\State Initialize $x^0=(x_i)_{i\in\mathcal{N}_u}\in\mathbb{R}^{2 N_u}$. Set $k= 0$.
	\Repeat
	\State For all $i\in\mathcal{N}_u$, solve $x_i^{k+1}=\hat{x}_i(x^k)$
	\State Exchange $x_i^{k+1}$ with neighbours $j\in\mathcal{N}^i,\;\forall i\in\mathcal{N}_u$. \label{eq-alg:exchange}
	\State Set $k\leftarrow k+1$
	\Until stopping criteria is satisfied
\end{algorithmic}
\caption{Network Nodes Self-positioning Algorithm}
\label{alg:scutari}
\end{algorithm}

%
\section{Underwater Navigation}
\label{sec:navigation-aid}
%
Underwater navigation is useful in applications where robotic devices or UVVs require orientation to perform some task, such as collect data from a sink node, or explore the seabed.
Many localization algorithms have been developed in the literature, both distributed and centralized, such as \cite{Costa2006,Lin2011,Tarrio2008,Savic2014}.
For the specific application of underwater navigation, the UVV needs to estimate its own position by measuring distances to every node within reach.
Since we assume the medium is isotropic, triangulation methods for outdoor localization are sufficient.

We assume that the network nodes know their position from a previous calibration, such as the one describe in Section~\ref{sec:network-positions}.
Then, these nodes transmit a signal to the UVV at fixed power and their local position (if unknown by the UVV).
The UVV estimates the distance to the specific node using the channel form that was described in Section~\ref{sec:network-description}.
In particular,
\begin{equation}
	\hat{d_i} = \frac{1}{a}({P_{tx}-P_{rx}}-b)
	\vspace{-0.5em}
\end{equation}
where $P_{tx}$ corresponds to the prefixed transmit power of the signal (in dBm); $P_{rx}$ the received power at the UVV (in dBm); and $a\approx-8.5~\rm{dB/m}$ and $b\approx -54.85~\rm{dB}$, as we previously derived.
Since the UVV measures all distance estimates directly, a centralized algorithm is preferable.
Moreover, this approach reduces the amount of data that needs to be transmitted through the channel.

Regarding the localization problem, the following problem is frequently proposed:
\begin{equation}
	\min_{u} \sum_{i=1}^n (\hat{d}_i^2-\Vert x_i -u \Vert^2)^2
	\label{eq:loc-problem}
	\vspace{-0.5em}
\end{equation}
where $x_i$ represent the network node's coordinates; $u$ the unknown target coordinates; $\hat{d}_i$ the estimated distances from the sensors to the target; and $n$ represents the number of measurements available.
Problem~\eqref{eq:loc-problem} is non-convex and is also suboptimal in the ML sense.
Nonetheless, because the noise variance is relatively small compared to the channel attenuation (as shown in \eqref{eq:noice-variance}), the solution to the problem yields a good estimate close to the theoretical bound of the ML estimator and gives an accurate estimate for underwater networks.

The ML estimator, which is discussed in \cite[Sec.II-A]{Beck2008}, is a more difficult problem.
The algorithms that suboptimally solve it may yield poorer estimates than the solution of~\eqref{eq:loc-problem}, so we did not consider this method in our simulations.

There are several algorithms in the literature that try to solve~\eqref{eq:loc-problem}.
For instance, the best linear unbiased estimator proposed in~\cite{Lin2011} yields a suboptimal result since it solves a relaxed approximation of~\eqref{eq:loc-problem}.
Other linear estimators, such as the one proposed in~\cite{Tarrio2008}, minimize an error function derived from~\eqref{eq:loc-problem} rather than the problem itself.
And majorization methods, such as the one proposed in~\cite{Costa2006}, solve the problem by successive convex approximations which converge to local minimization points.

An optimal solution of~\eqref{eq:loc-problem} is however available as presented in~\cite{Beck2008}.
Its solution can be found by reformulating the objective to a non-convex quadratic problem with a single quadratic constraint, which presents strong duality and can be solved optimally \cite{More1993}.

We skip the derivation as it can be followed in~\cite[Sec.II-B]{Beck2008} and directly illustrate the equivalent problem:
\begin{IEEEeqnarray}{c}
	\begin{IEEEeqnarraybox}[][c]{r'l}
		\underset{\lambda\in \mathbb{R}}{\rm{solve}}&  \hat{y}(\lambda)^T D y(\lambda) + 2f^T \hat{y}(\lambda)=0 \\
		\text{s.t.} &  \hat{y}(\lambda) = ( B^TB+\lambda D)^{-1}(B^Tc-\lambda f) \\
		& B^TB+\lambda D \succeq 0  
	\end{IEEEeqnarraybox}
	\label{eq:equiv-pos}
\end{IEEEeqnarray}
where $\hat{y}$ is an estimator of $y^T=[u^T,\alpha]\in\mathbb{R}^3$ and
\begin{IEEEeqnarray}{c}
	\begin{IEEEeqnarraybox}[][c]{l'l}
		B = 
		\begin{pmatrix}
		 	-2 x_1^T & 1 \\
		 	\vdots & \vdots
		 	-2 x_n^T & 1
		\end{pmatrix}, &
		c = \begin{pmatrix}
			\hat{d}_1^2-\Vert x_1\Vert^2 \\
			\vdots \\
			\hat{d}_n^2-\Vert x_n\Vert^2
		\end{pmatrix}
		\\
		D = 
		\begin{pmatrix}
		 	\mathbf{I}_{n} & \mathbf{0}_{n\times 1} \\
		 	\mathbf{0}_{1\times n} & 0
		\end{pmatrix}, &
		f = \begin{pmatrix}
			\mathbf{0}_{n\times 1} \\
			-0.5
		\end{pmatrix}.
	\end{IEEEeqnarraybox}
\end{IEEEeqnarray}

The objective function of~\eqref{eq:equiv-pos} is monotone in $\lambda$, so a bisection algorithm can be applied and converges very fast to the unique root of the system~\cite{Beck2008,More1993}.
The specific steps are described in Algorithm~\ref{alg:stoica}.

\begin{algorithm}
\begin{algorithmic}[1]
	\State Initialize $\underline{\lambda}=-{1}/{\max \rm{eig}(D,B^TB)}$, $\overline{\lambda} = 100$.
	\While{$\hat{y}(\lambda)^T D y(\lambda) + 2f^T \hat{y}(\lambda) >= 0$}
		$\overline{\lambda} \leftarrow 2 \overline{\lambda}$
	\EndWhile
	\State Set $\lambda \leftarrow \frac{1}{2}(\overline{\lambda}+\underline{\lambda})$.
	\While{ $\overline{\lambda}-\underline{\lambda} \geq \epsilon$ }
	\If{$\hat{y}(\lambda)^T D y(\lambda) + 2f^T \hat{y}(\lambda) >= 0$}
		$\underline{\lambda} \leftarrow \lambda$
	\Else\,
		$\overline{\lambda} \leftarrow \lambda$.
	\EndIf
	\EndWhile
	\State Set $[u^T,\alpha]\leftarrow \hat{y}^T(\lambda)$. \textbf{Return} $u$.
\end{algorithmic}
\caption{Bisection algorithm for UVV localization}
\label{alg:stoica}
\end{algorithm}

%
\section{Simulations}
\label{sec:simulations}
%
For the simulations we first analyze the results presented in Section~\ref{sec:network-positions}.
We consider a network of 27 nodes forming a grid, where $|\mathcal{N}_a|=4$ and $|\mathcal{N}_u|=23$. 
The nodes have a separation of 5 meters in the horizontal and vertical directions as depicted in Figure~\ref{fig:ms_network}.
We also considered that two nodes are neighbours if they are within 10 meters of each other.
We used the channel coefficients derived in Section~\ref{sec:network-description}, where the channel model is given by~\eqref{eq:linear-model}.
Finally, we have considered a deviation noise value of 0.63~m as a trade-off between accuracy and energy consumption in the estimation procedure.

In Figure~\ref{fig:ms_network} we represent the result of Algorithm~\ref{alg:scutari} after 50 iterations in the case where the network does not miss any packet. 
In Figure~\ref{fig:ms_distributed} we represent the absolute mean error of all nodes with unknown position versus the number of iterations (message exchanges).
We plot several curves with different probabilities of packet loss, i.e., probabilities of 0\%, 5\%, 10\% and 20\%.
If a packet is lost, the last known message of the node who could not send the packet correctly is used.
Specifically, the last known estimate of the node is used for the computation of step~\ref{eq-alg:exchange}, from Algorithm~\ref{alg:scutari}.
Figure~\ref{fig:ms_distributed} shows that the algorithm is robust against packet losses, and the effect that can be observed in this kind of scenario is a slower convergence rate.

Next, we consider the tracking method proposed in Section~\ref{sec:navigation-aid}.
We plot a qualitative localization result of a target moving in the previous network, which is depicted in Figure~\ref{fig:tracking}.
The network is the same as in the previous case considering $|\mathcal{N}|=27$.
We also considered that the target node can measure distances from nodes within 8 meters.
Note also that Algorithm~\ref{alg:stoica} solves problem~\eqref{eq:loc-problem} in an optimal manner, although the problem is non-convex.
The mean absolute error of all points is about 0.75~m.

\begin{figure}
	\centering
%
%
\begin{tikzpicture}

\begin{axis}[%
width=\figurewidthB,
height=\figureheightB,
at={(0.758in,0.481in)},
scale only axis,
xmin=-25,
xmax=25,
xtick={-50, -40, -30, -20, -10,   0,  10,  20,  30,  40,  50},
xmajorgrids,
ymin=-10,
ymax=10,
ytick={-10, -5, 0,   5,   10},
ymajorgrids,
axis background/.style={fill=white},
legend style={at={(1,0)},anchor=south east,legend cell align=left,align=left,draw=white!15!black,font=\tiny}
]
\addplot [color=black,only marks,mark=o,mark options={solid}]
  table[row sep=crcr]{%
-20	5\\
20	5\\
-20	-5\\
20	-5\\
};
\addlegendentry{Anchor nodes};

\addplot [color=blue,only marks,mark=asterisk,mark options={solid}]
  table[row sep=crcr]{%
5	-5\\
-15	5\\
-10	5\\
-5	5\\
0	5\\
5	5\\
10	5\\
15	5\\
10	-5\\
-20	0\\
-15	0\\
-10	0\\
-5	0\\
0	0\\
5	0\\
10	0\\
15	0\\
20	0\\
15	-5\\
-15	-5\\
-10	-5\\
-5	-5\\
0	-5\\
};
\addlegendentry{Unknown nodes};

\addplot [color=red,only marks,mark=+,mark options={solid}]
  table[row sep=crcr]{%
4.95717441873076	-4.70235373075687\\
-14.2988883374875	4.75076029960464\\
-9.83931827231618	4.52451879539379\\
-5.59398161219919	4.58362336130893\\
-0.577524600601488	4.65674101033378\\
4.38923627463899	5.18154959167655\\
9.3687850614341	5.49734716334456\\
14.3572676916019	5.62795125836792\\
9.5782314155997	-4.37097328649032\\
-19.9327022733426	0.151672886022238\\
-14.9334765819032	-0.417740626581645\\
-9.67794277064283	-0.489312173095569\\
-4.89714164967931	-0.561183524521104\\
-0.292472551164496	-0.372102485135133\\
5.07797436079883	0.5135545640701\\
9.83629129602996	0.960807508045824\\
15.0521496681986	0.162529274914121\\
19.7212585730319	0.476172001060306\\
14.4141601140967	-4.78301731460413\\
-14.3976389362063	-5.41209701196768\\
-10.0367486558379	-5.39077022455523\\
-4.8562861179838	-5.47159969988642\\
0.236499002468726	-5.14905074052705\\
};
\addlegendentry{Estimates};

\end{axis}
\end{tikzpicture}%
	\caption{Network self-localization qualitative result after 50 iterations of Algorithm~\ref{alg:scutari} with 0\% probability of packet loss in neighbour transmissions.}
	\label{fig:ms_network}
\end{figure}
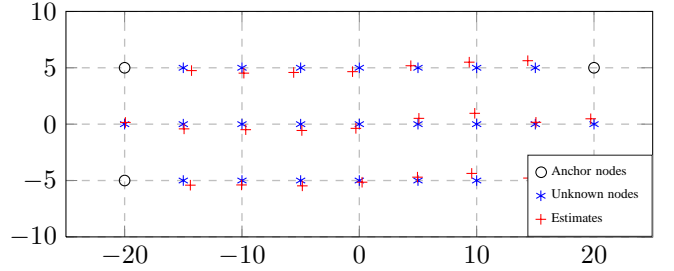
\begin{figure}
	\centering
    \input{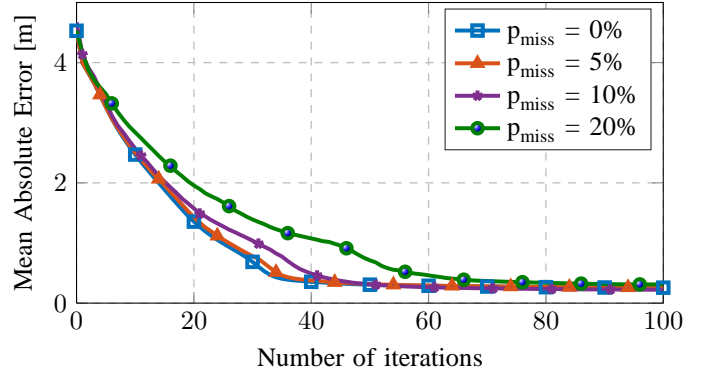}
	\caption{Network self-localization mean absolute error results versus the number of iterations of Algorithm~\ref{alg:scutari}. The different curves indicate the probability of packet loss of some node to all of its neighbours.}
	\label{fig:ms_distributed}
\end{figure}
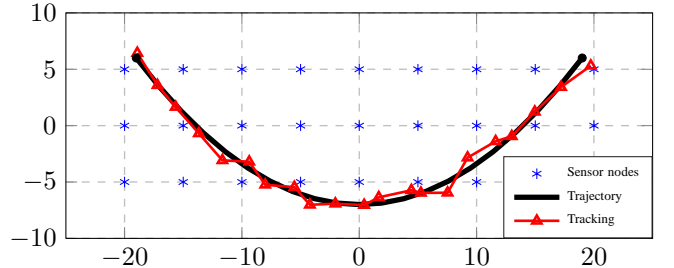
\begin{figure}
	\centering
%
%
\begin{tikzpicture}

\begin{axis}[%
width=\figurewidthB,
height=\figureheightB,
at={(0.886in,0.462in)},
scale only axis,
xmin=-25,
xmax=25,
xtick={-50, -40, -30, -20, -10,   0,  10,  20,  30,  40,  50},
xmajorgrids,
ymin=-10,
ymax=10,
ytick={-10, -5, 0,   5,   10},
ymajorgrids,
axis background/.style={fill=white},
legend style={at={(1,0)},anchor=south east,legend cell align=left,align=left,draw=white!15!black,font=\tiny}
]
\addplot [color=blue,only marks,mark=asterisk,mark options={solid}]
  table[row sep=crcr]{%
-20	5\\
-15	5\\
-10	5\\
-5	5\\
0	5\\
5	5\\
10	5\\
15	5\\
20	5\\
-20	0\\
-15	0\\
-10	0\\
-5	0\\
0	0\\
5	0\\
10	0\\
15	0\\
20	0\\
-20	-5\\
-15	-5\\
-10	-5\\
-5	-5\\
0	-5\\
5	-5\\
10	-5\\
15	-5\\
20	-5\\
};
\addlegendentry{Sensor nodes};

\addplot [color=black,solid,line width=2.0pt]
  table[row sep=crcr]{%
19	6\\
17.1	3.53\\
15.2	1.32\\
13.3	-0.629999999999999\\
11.4	-2.32\\
9.5	-3.75\\
7.6	-4.92\\
5.7	-5.83\\
3.8	-6.48\\
1.9	-6.87\\
0	-7\\
-1.9	-6.87\\
-3.8	-6.48\\
-5.7	-5.83\\
-7.6	-4.92\\
-9.5	-3.75\\
-11.4	-2.32\\
-13.3	-0.630000000000002\\
-15.2	1.32\\
-17.1	3.53\\
-19	6\\
};
\addlegendentry{Trajectory};

\addplot [color=red,solid,line width=1.0pt,mark=triangle,mark options={solid}]
  table[row sep=crcr]{%
19.7315713729208	5.31359506882931\\
17.2009185438887	3.41839254381405\\
14.9582087777882	1.22782760047571\\
12.9966734192422	-0.942249379246128\\
11.6232661636198	-1.39204614762421\\
9.2321626652141	-2.83981864849555\\
7.52774472406454	-5.95112901819737\\
5.25383114133915	-5.96372979510521\\
4.44133310412328	-5.73160020123911\\
1.66540178788737	-6.35898943124119\\
0.417127812473174	-7.04551001864673\\
-2.03589166263663	-6.92205990377284\\
-4.20593644936646	-7.01913156597768\\
-5.52779108628739	-5.45525652028588\\
-8.01378217456136	-5.23989267416159\\
-9.37692606354015	-3.19961761726295\\
-11.6622282574448	-3.07612950956501\\
-13.6697712662961	-0.686232702989685\\
-15.6472407980832	1.65705010281941\\
-17.1937454647254	3.58617299080588\\
-18.9149394175778	6.44272217255132\\
};
\addlegendentry{Tracking};

\addplot [color=black,mark size=1.5pt,only marks,mark=*,mark options={solid,fill=black},forget plot]
  table[row sep=crcr]{%
19	6\\
};
\addplot [color=black,mark size=1.5pt,only marks,mark=*,mark options={solid,fill=black},forget plot]
  table[row sep=crcr]{%
-19	6\\
};
\end{axis}
\end{tikzpicture}%
	\caption{Tracking performance taking into account 8~m of radio sensitivity.}
	\label{fig:tracking}
	\vspace{-1em}
\end{figure}

%
\section{Conclusion}
%
We studied and proposed algorithms for localization techniques in an underwater sensor network.
We considered the specific underwater channel measurements of~\cite{Jimenez2016,Mena2016}.
We took into account packet loss considerations due to channel changes and also low connectivity between nodes, as well as low data rate packet limits.
Finally, Algorithm~\ref{alg:scutari} can be solved in a distributed manner and Algorithm~\ref{alg:stoica} is solved optimally.

%
\section*{Acknowledgment}
%
The authors would like to thank Juan Domingo Santana Urbin (ULPGC) for all the help provided to make the underwater measurements.
Also, the authors would like to thank Gabriel Juanes and Raul Santana (PLOCAN) for setting up the measurement testbed in the pier and into the sea.

\vfill
%
\bibliographystyle{IEEEtran}
\bibliography{refs}
%

\end{document}